\documentstyle[11pt,paspconf]{article}
\begin{document}

\title{Bayesian Approach to Foreground Removal}
\author{J. Jewell\altaffilmark{1}}
\affil{Department of Astronomy and Astrophysics, University of Chicago,
5640 S. Ellis Ave., Chicago, IL  60637}
\author{C. R. Lawrence and S. Levin}
\affil{Jet Propulsion Laboratory, MS 169-506, 4800 Oak Grove Drive,
Pasadena, CA 91109-8099}

\altaffiltext{1}{Email: jewell@oddjob.uchicago.edu}

\begin{abstract}
Our ability to extract the maximal amount of information from 
future observations at gigahertz frequencies depends on our ability
to separate the underlying cosmic microwave background (CMB)
from galactic and extragalactic foregrounds.
We review the separation problem and its formulation
within Bayesian inference, give examples of specific solutions
with particular choices of prior density, and finally
comment on the generalization of Bayesian methods to a multi-resolution
framework.  We propose a strategy for the regularization of solutions
allowing a spatially varying spectral index, and discuss possible
computational approaches such as multi-scale stochastic relaxation. \\
\\
{\bf Key Words} methods: data analysis - techniques: image analysis - cosmic
mcrowave background
\end{abstract}

\section{Introduction}
Future observations from ground based, balloon borne, and
satellite missions at gigahertz frequencies
will contain a wealth of information.  Our ability to extract the
maximal amount of information from these
experiments depends on our ability to separate the underlying
cosmic microwave background (CMB) from galactic and extragalactic foregrounds.
Anticipated components of the total foreground emission include
synchrotron, free-free, and dust emission in our
own galaxy, plus various extragalactic point sources.
The interesting cosmological information, relevant for testing
inflation and determining cosmological parameters, is found
at sub-degree angular scales, but it is at these angular scales
that we begin to resolve the microphysics of the galaxy,
potentially complicating the separation of the underlying cosmic
and foreground signals.

It has been demonstrated that, for foregrounds simulated
by extrapolating  existing observations to the relevant frequencies 
and spatial resolutions of the next generation of experiments, 
an accurate reconstruction of the CMB is possible (Brandt  et al. 1994,
Tegmark \& Efstatiou 1996, Hobson et al. 1998).
However, these foreground scenarios involve
overly simple assumptions about the emissivity or spatial structure of the
foregrounds. These methods assume the foregrounds have a power law 
emissivity, $I_{\nu} \propto \nu^{\alpha}$, and 
that the spectral index $\alpha$ is known or can be adaquately inferred
from the emission averaged over some spatial scale.
The spatial structure of the variation of the spectral
index, at the frequencies relevant for observations of the CMB,
is in fact {\it a priori} unknown. 
Although further complexities in foreground models are a  nuisance
with respect to CMB extraction, they
represent  a ``gold mine'' of interesting physical quantities
for ISM studies (A. Lazarian, 1998).

Because of unknown complexities in the physics of the foregrounds,
we have a family of models, in which previous assumptions,
such as a spatially constant spectral index, are 
successively relaxed.
Inference within models with increasingly many
degrees of freedom  becomes impossible (completely degenerate),
so the ability to constrain solutions is necessary for the extraction
of useful information from the data.  Bayesian inference
provides a unifying framework within which the separation problem can
be addressed in its full range of complexity.
In addition, the Bayesian framework can provide computational solutions for more
complicated, yet physically motivated, models through algorithms
including stochastic relaxation.

In this article, we will review the separation problem and its formulation
within Bayesian inference, give examples of specific solutions
with particular choices of prior density, and finally
comment on the generalization of Bayesian methods to a multi-resolution
framework.  The paper is meant to convey ideas of variations on
current analysis strategies and present the range of
possible approaches from linear filtering to mutli-scale stochastic
relaxation.

\section{Review of the Separation Problem}
The anticipated galactic foregrounds contributing to
the total detected emission at frequencies of interest for CMB observations
include free-free, synchrotron, and dust.  The total specific intensity
integrated along the line of sight from these sources is
(following the convention in (Tegmark \& Efstatiou 1996))
\begin{equation}
I(r,\nu) = \begin{array}{l}
\left[  (\mbox{$270.2$ ${\rm MJy}$ ${\rm sr}^{-1}$})  \left(
\frac{x_{0}^{3}}{e^{x_{0}}-1} \right) \right. \\
\left. + (\mbox{$24.8$ ${\rm MJy}$ ${\rm sr}^{-1}$ ${K}^{-1}$})
 \left( \frac{x_{0}^{2}}{\sinh (x_{0}/2)} \right)^{2} 
\delta T^{(CMB)} (r) \right] \\
+  I^{(ff)}_{0}(r)\left( \frac{\nu}{\nu_{0}} \right)^{\alpha_{ff}}
+  I^{(s)}_{0} (r) \left( \frac{\nu}{\nu_{0}} \right)^{\alpha_{s}} \\
+ I^{(dust)}_{0}(r) \left( \frac{\nu}{\nu_{0}} \right)^{2+ \beta(r)}
\left(  \frac{x_{d}^{3}}{e^{x_{d}}-1} \right)
\end{array}
\end{equation}
where $I_{0}$ is a dimensionless map at a convenient reference frequency
for the specific physical component,
$x_{d} = h \nu / k T_{d} \sim \nu / 417 {\rm GHz}$ for a dust temperature
of $T_{d}=20 {\rm K}$, $T_{0}$ is the temperature of the CMB blackbody, 
and $x_{0} = \nu / 56.8 {\rm GHz}$.
Subtracting the isotropic blackbody term and converting to brightness
temperature (in micro-Kelvin) gives
\begin{equation}
T(r,\nu) = \delta T^{(CMB)}(r) + \sum_{i} A^{(i)}(\nu) T^{(i)}_{0}(r)
\left( \frac{\nu}{\nu_{0}} \right)^{\beta^{(i)}(r)}
\end{equation}
where $A^{(i)}(\nu)$ is the spatially independent frequency dependence
for the $i^{th}$ physical component, with spatial variations
$\beta^{(i)}(r)$.  Note that the CMB 
is a frequency-independent, zero-mean field in these units.  The frequency
dependence of free-free emission is given by
physics assuming typical electron densities
and temperatures, while the synchrotron and dust spectral
indices can be spatially varying due to spatial dependence of the
galactic magnetic field and dust grain properties.
The data returned from an experiment will be the integrated brightness
temperature over the bandpass of the frequency channels of the instrument
(with center frequency $\nu_{i}$) with additive noise $\eta(r,\nu_{i})$ giving
\begin{equation}
T_{obs}(r,\nu_{i}) = \delta T^{(CMB)}(r) + 
\sum_{j} A^{(j)}(\nu_{i}) T^{(j)}_{0}(r) \left(
\frac{\nu_{i}}{\nu_{0}} \right)^{\beta^{(j)}(r)} + \eta(r,\nu_{i})
\end{equation}
(note that the maps at each frequency are the result of an initial
processing stage from the time series returned from the experiment excuting
a particular scan strategy.  We will not include details
of this stage of analysis here.)

\section{Bayesian Formulation}
Given observations $T_{obs}$ at the frequency channels of the instrument,
we would like to recover the amplitude and spectral index
for each of the assumed present physical components.
The probability $p[T_{obs}|T^{(i)}_{0},\beta^{(i)}]$
of the data given the underlying amplitude $T^{(i)}_{0}(r)$
and spectral indices $\beta^{(i)}(r)$ is given by the statistics of the noise process
and scan strategy.  This probability is known as the likelihood, where
$\log p[T_{obs}|T^{(i)}_{0},\beta^{(i)}] \sim - \chi^{2}$ up to an additive constant
(the normalization constant), and for pixel independent Gaussian noise
\begin{equation}
\chi^{2}  = 
\left[ T_{obs} - \sum_{j} A^{(j)} T^{(j)} \left(
\frac{\nu_{i}}{\nu_{0}} \right)^{\beta^{(j)}} \right]^{T} N^{-1}
\left[ T_{obs} - \sum_{j} A^{(j)} T^{(j)} \left(
\frac{\nu_{i}}{\nu_{0}} \right)^{\beta^{(j)}} \right]
\end{equation}
with $N_{rs} = \langle \eta^{2} \rangle \delta_{rs}$.

To find the underlying variables $(T^{(i)}_{0}(r),\beta^{(i)}(r))$,
we could simply try to minimize $\chi^{2}$.  In the limit that
we have an overdetermined system with high signal-to-noise ratio,
we can obtain good results with linear methods such as
singular value decomposition (with the spectral index assumed known),
or a non-linear $\chi^{2}$ method such as used by (Brandt et al. 1994).
However, as noted in (Brandt et al. 1994), the recovery of the
amplitude and spectral index
for all the physical components in the presence of noise is numerically
unstable.   Some means of regularizing the
solution is needed.  

There are several stategies that can be employed to
regularize solutions.  (Brandt et al. 1994) discuss two reasonable
simplifications - either assume the contribution of a particular
physical component is negligible at specific frequency channels,
or find the spectral index from the average emission on patches
of sky $10^{\circ} \times 10^{\circ}$, and then find the amplitudes
at all pixels holding the spectral index fixed within the given patch
of sky.  

Wiener filtering is another strategy that has been 
demonstrated successfully on simulated
foregrounds, and is approporiate in cases where we have information
about the spatial power spectrum of the foregrounds.
The original method of (Tegmark \& Efstatiou 1996) assumes the
foregrounds are uncorrelated, with a power spectrum of the form
$C_{l} \propto l^{-3}$.  The Wiener matrix $W$
is constructed so that $T^{(i)}_{0} = WT_{obs}$, and
the expected value of the residual error is a minimum, giving
the solution
\begin{equation}
T^{(i)}_{0}= \left[ A^{T}N^{-1}A +C^{-1} \right]^{-1} A^{T}N^{-1}T_{obs}
\end{equation}
where $W=\left[ A^{T}N^{-1}A +C^{-1} \right]^{-1} A^{T}N^{-1}$,
$A$ is the frequency response matrix as above, and
$C_{ij}= \langle T^{(i)}_{0}(r_{i}) T^{(i)}_{0}(r_{j}) \rangle$ is the power
spectrum of the $i^{th}$ foreground component.

As observed by (Hobson et al. 1998), solutions to the foreground separation
problem can be generically formulated within Bayesian inference.
In Bayesian inference, the posterior probability is interpreted
as the figure of merit of a solution,
quantifying  our degree of confidence, and given by
\begin{equation}
p[T^{(i)}_{0},\beta^{(i)}|T_{obs}] \propto
p[T_{obs}|T^{(i)}_{0},\beta^{(i)}] p[T^{(i)}_{0},\beta^{(i)}]
\end{equation}
The first term on the right-hand side is the likelihood as discussed
above.  The term $p[T^{(i)},\beta^{(i)}]$ is the prior density
which effectively regularizes solutions through a 
statistical characterization of the foregrounds known or assumed 
{\it a priori}.  Choosing a Gaussian prior with the assumed power spectrum 
of the foregrounds gives the Wiener filter solution as the maximum
posterior solution (as pointed out by (Hobson et al. 1998)).
The maximum entropy (MAXENT) method of (Hobson et al. 1998))
finds the maximum posterior solution with the log-prior given by
the entropy.  The method of (Brandt et al. 1994) effectively uses a
uniform prior over an allowed interval for the fluctuations within a simplified
model.

We also note a slight variation on the Wiener filtering solution, in which
we include previous observations as a spatial template for
substraction.   For example, denoting the previous observations
extrapolated to the frequencies and resolution of
interest $T_{other}$, we can use the prior (in matrix notation)
\begin{equation}
- \log p[T^{(i)}_{0}|T_{other}] \sim \sum_{i}
(T^{(i)}_{0})^{T} C^{-1} (T^{(i)}_{0})
+ \sum_{i} (T^{(i)}_{0}) \Lambda_{i} (T_{other})
\end{equation}
where $\Lambda$ quantifies the relative weight of the coupling of
$(T_{other})$ to inferences made about $(T^{(i)}_{0})$.
This prior gives the posterior
\begin{equation}
- \log p[T^{(i)}_{0}|\beta^{(i)},T_{obs},T_{other}] \sim \chi^{2} +
(T^{(i)}_{0})^{T} C^{-1} (T^{(i)}_{0}) + (T^{(i)}_{0}) \Lambda (T_{other})
\end{equation}
The solution which maximizes the posterior is then
\begin{equation}
T^{(i)}_{0}(r) = \left[ A^{T}N^{-1}A +C^{-1} \right]^{-1}
\left[A^{T}N^{-1}T_{obs}-\Lambda T_{other} \right]
\end{equation}
Coupling to previous observations as a spatial template,
and choosing the maximum
posterior solution, gives a subtraction of a particular
foreground component through the filtered map $\Lambda T_{other}$.
This method has the obvious danger of
the false addition of correlation in the various components, however we might
want to include information from previous observations at large angular
scales, where instrumental noise and systematic effects might be
typically very low.  

\section{Separation Problem in a Multi-Resolution Setting}
There is a natural way to define fluctuation maps at various scales
by adopting a multi-resolution approach (which will not be discussed in
great detail here - we follow the discussion in (Langer et al. 1993)).  
A multi-resolution approach can be implemented with 
a smoothing function to recursively generate
low-pass, or Gaussian, images and band-pass, or
Laplacian, images according to
\begin{equation}
\begin{array}{l}
T^{(i)}_{0}(r,\sigma_{j}) = G \ast T^{(i)}_{0}(r,\sigma_{j-1}) \\
\\
LT^{(i)}_{0}(r,\sigma_{j-1}) = L \ast T^{(i)}_{0}(r,\sigma_{j-1})
\end{array}
\end{equation}
where $L=I-G$.  An example for the smoothing filter is the discrete
approximation to a Gaussian provided by a separable filter
$G(i,j)=g_{i}g_{j}$, where for example we can take
$g_{i} = [ 1/16, 1/4, 3/8, 1/4, 1/16]$.
Note that this recursive scheme gives a non-orthogonal and overcomplete wavelet
decomposition.  The importance of a non-orthogonal and overcomplete basis
for image analysis (as opposed to image compression) 
has been stressed elsewhere
(see (Langer et al. 1993) and references therein for a discussion
of this point).  The basic motivation for a non-orthogonal, overcomplete
basis is that the coefficients in an
orthogonal wavelet bases can become diffuse upon translations
of the original image.

In such a scale-space decomposition, we have fluctuation maps at various
scales
\begin{equation}
LT(r,\nu,\sigma) = L \delta T^{(CMB)}(r,\sigma) + 
\sum_{i} A^{(i)}(\nu) LT^{(i)}(r,\nu,\sigma)
\end{equation}
where we have defined the effective scale-space spectral index as
\begin{equation}
T^{(i)}(r,\nu,\sigma) = T^{(i)}_{0}(r,\sigma) \left(
\frac{\nu}{\nu_{0}} \right)^{\beta^{(i)}(r,\sigma)}
\end{equation}
Note that the above implicitly defines the spectral index in terms of the
difference in the log brightness temperature at spatial scale
$\sigma$ of a physical component at two frequencies.  

The motivation for the above conventions is simply that, given the
spectral indices for the physical components at a given scale, we
can iteratively update the fluctuation maps (the simple separation
problem traditionally considered).  However, when adjusting the
spectral index, we need to work with the total brightness temperature
at the next scale, since
\begin{equation}
\begin{array}{l}
c_{\nu} [\beta^{(i)}(r,\sigma_{j-1}) - \beta^{(i)}(r,\sigma_{j}) ]=
 \log \left[ T^{(i)}(r,\nu,\sigma_{j})
+ LT^{(i)}(r,\nu,\sigma_{j-1}) \right] \\
\\
- \log \left[( T^{(i)}_{0}(r,\sigma_{j})
+ LT^{(i)}_{0}(r,\sigma_{j-1}) )
\left( \frac{\nu}{\nu_{0}} \right)^{\beta^{(i)}(r,\sigma_{j})} \right]
\end{array}
\end{equation}
where $c_{\nu} = \log (\nu / \nu_{0})$.
The scale-space data can then be written as
\begin{equation}
T_{obs}(r,\nu_{i},\sigma) = \delta T^{(CMB)}(r,\sigma) + 
\sum_{j} A^{(j)}(\nu_{i}) T^{(j)}_{0}(r,\sigma) \left(
\frac{\nu_{i}}{\nu_{0}} \right)^{\beta^{(j)}(r,\sigma)} 
+ \eta(r,\nu_{i},\sigma)
\end{equation}
where the noise $\eta(r,\nu_{i},\sigma)$ has covariance
matrix $\langle (\eta^{T} G^{T}G \eta) \rangle^{-1}$.

We explored a simple separation problem (standard galactic foregrounds
with assumed known and spatially constant spectral index) within the above
multi-resolution context as a test case in anticipation of more complex
foreground scenarios.  We considered only a single Gaussian
and Laplacian component, so that
\begin{equation}
T^{(i)}_{0}(r,0) = LT^{(i)}_{0}(r,0) + T^{(i)}_{0}(r,\sigma)
\end{equation}
For pixel independent Gaussian noise, the majority of the 
noise power is contained in the
Laplacian component of the total observed emission, and we
expect the foregrounds to be locally smooth with sparsely distributed
dominant features in the map.  We are therefore motivated
to assume a quadratic prior in the Laplacian component
\begin{equation}
- \log p[LT^{(i)}_{0}] \sim  \theta^{i} \sum_{j} (LT^{(i)}_{0}(r_{j}))^{2}
\end{equation}
where $\theta^{j}$ is a parameter quantifying the relative weight
between the prior and likelihood (see the review in (Geman 1990)
for other applications of this type of prior).
The maximum posterior solution is given by,
\begin{equation}
T^{(i)}_{0}=[A^{T}N^{-1}A+B-H]^{-1}A^{T}N^{-1} T_{obs}
\end{equation}
where the matrix elements $B^{j}_{rs}=\theta^{j}\delta_{rs}$, and
$H^{j}_{rs}= \theta^{j}[2G_{rs}-(GG)_{rs}]$.

In order to implement the above however, an estimate for $\theta^{(j)}$
was needed.  We first found a preliminary solution with $\chi^{2}$
minimization, equivalent to the above solution
with $\theta^{j}=0$.  After this solution was obtained, we
estimate $\theta^{j}$ from the initial solution as
\begin{equation}
\theta^{j}  =  \frac{1}{2 \langle (LT^{(j)}_{0})^{2} \rangle} 
\end{equation}
where the angle brackets denote the spatial average over
the initial solution.
After fixing $\theta^{j}$, we continue with an iterative improvement
on the Laplacian component according to
\begin{equation}
\label{eq:Lapit}
\begin{array}{lll}
LT^{(i)}_{0}(n+1) & = & \alpha A^{T}N^{-1}[T_{obs}-A GT^{(i)}_{0}(n)] + 
MLT^{(i)}_{0}(n) \\
T^{(i)}_{0}(n+1) & = & T^{(i)}_{0}(n) + LT^{(i)}_{0}(n+1) - 
LT^{(i)}_{0}(n) \\
GT^{(i)}_{0}(n+1) & = & G \ast T^{(i)}_{0}(n+1)
\end{array}
\end{equation}
where the iteration matrix is $M=I-\alpha(A^{T}N^{-1}A + B)$
and $\alpha^{k}=1/(A^{T}N^{-1}A+B)_{kk}$.
The iteration relaxes to the desired maximum posterior solution.
The motivation for this iteration is that, to a good
approximation, the low-pass filtered $\chi^{2}$ solution
is very close to the true low-pass filtered map (since the low-pass
noise rms is drastically reduced).  The needed improvement
to the solution is in the Laplacian component, leading us to the
iteration above.

\section{Parameter Estimation}
The approximation for $\theta^{(j)}$ gave good results for our specific
simple test case above, but in general we will want a more consistent
approach to fixing parameters.  Within Bayesian inference, the formal
way to treat parameters is to include them in the posterior
as another piece of information to be drawn from the data.  In general,
for a collection of weight parameters in the prior (such as the
parameters $\theta^{(j)}$ in our simple test case above), the
full posterior becomes
\begin{equation}
p[T^{(i)}_{0},\beta^{(i)}, {\bf \theta}|T_{obs}] \propto
p[T_{obs}|T^{(i)}_{0},\beta^{(i)}] p[T^{(i)}_{0},\beta^{(i)}|{\bf \theta}]
p[{\bf \theta}]
\end{equation}
where ${\bf \theta}$ is the vector of weight parameters.  Note that
the likelihood is not dependent on the parameters in the prior.  The parameters
can then be chosen from the density $p[\theta, T_{obs}]$,
obtained by marginalization over $(T^{(i)}_{0},\beta^{(i)})$, so that
\begin{equation}
p[{\bf \theta},T_{obs}] \propto p[{\bf \theta}] \int d[T^{(i)}_{0},\beta^{(i)}]
p[T_{obs}|T^{(i)}_{0},\beta^{(i)}] p[T^{(i)}_{0},\beta^{(i)}|{\bf \theta}]
\end{equation}
The density $p[\theta, T_{obs}]$ is typically a very sharply
peaked function.  A reasonable choice of parameter values is then
the one that maximizes the likelihood
\begin{equation}
\theta = {\rm max}_{\theta} p[T_{obs}|\theta]
\end{equation}
This is the strategy employed by (Hobson et al. 1998) to
fix the relative weight between an entropic prior and the likelihood.

\section{Generalization of Multi-Resolution Bayesian Inference}
Returning to the multi-resolution setting described before, we
want to discuss Bayesian methods proceeding from coarse
to fine scales.  For notational purposes, let
$T^{(i,j)}_{0}(r)= T^{(i)}_{0}(r,\sigma_{j})$ be the emission at
the reference frequency of the $i^{th}$ phyiscal
component at the $j^{th}$ scale.  Let
$T^{(i,j)}_{m}(r) = T^{(i)}(r,\nu_{m},\sigma_{j})$ be the
emission in frequency channel $m$
of the $i^{th}$ physical component at the $j^{th}$ scale, and
finally let $\beta^{(i,j)}(r) = \beta^{(i)}(r,\sigma_{j})$
be similarly defined.
We can factor the probability density according to
\begin{equation}
p[T^{(i)}_{0},\beta^{(i)}|T_{obs}] \propto \left(
\prod_{\sigma_{j}} p[T^{(i,j)}_{0},\beta^{(i,j)}|
T^{(i,j+1)}_{0},\beta^{(i,j+1)},T_{obs}] \right)
\end{equation}
Each term above has the form
\begin{equation}
p[T^{(i,j)}_{0},\beta^{(i,j)}|
T^{(i,j+1)}_{0},\beta^{(i,j+1)},T_{obs}] \propto \begin{array}{l}
p[T_{obs}|T^{(i,j)}_{0},\beta^{(i,j)}] \\
\times  p[T^{(i,j)}_{0},\beta^{(i,j)}|T^{(i,j+1)}_{0},\beta^{(i,j+1)}]
\end{array}
\end{equation}
where the first term is just the generalization
of $\chi^{2}$ at scale $\sigma_{j}$, and the second term is
the prior.  We can factor the prior so that
\begin{equation}
p[T^{(i,j)}_{0},\beta^{(i,j)}|T^{(i,j+1)}_{0},\beta^{(i,j+1)}] = \begin{array}{l}
p[\beta^{(i,j)}|T^{(i,j)}_{0},T^{(i,j+1)}_{0},\beta^{(i,j+1)}] \\
\times p[T^{(i,j)}_{0}|T^{(i,j+1)}_{0},\beta^{(i,j+1)}]
\end{array}
\end{equation}
The last term $p[T^{(i,j)}_{0}|T^{(i,j+1)}_{0},\beta^{(i,j+1)}]$
is the prior on the amplitudes at the reference frequency with the
constraint that the low-pass filtered map of the finer scale
amplitude agree with the coarse scale above, 
$T^{(i,j+1)}_{0} = G \ast T^{(i,j)}_{0}$.  We could use
multi-resolution generalizations of the previous priors for this
term, such as multi-resolution MAXENT methods (Starck \& Pantin 1996,
Starck et al. 1998).
The term $p[\beta^{(i,j)}|T^{(i,j)}_{0},T^{(i,j+1)}_{0},\beta^{(i,j+1)}]$
is a prior on the spectral index given the amplitudes at the
reference frequency, and can be formulated in terms of
statistical characterizations of $T^{(i,j)}_{m}$, the emission of the physical
components at each frequency.

One possible strategy in constructing a prior for $\beta^{(i,j)}$
is motivated by the observation that the foregrounds
are non-Gaussian.  Realizations
of a non-Gaussian process can be more compressable in a suitable
basis (depending on the type of non-Gaussianity), 
so that fewer wavelet coefficients
are needed to reconstruct the image for a given rms error than
in the Fourier basis.  Therefore, we expect to be able to make
good predictions about $T^{(i,j)}_{m}$ from $T^{(i,j+1)}_{m}$
simply by assuming that there are no new features at the finer scale.
This approximation is a good one over large scale-space intervals
for dominant features.

Dominant features can be associated with wavelet maxima,
and tracking the location and amplitude of wavelet
maxima across scale provide a way to contrain proposed
fine scale maps $T^{(i,j)}_{m}$ given $T^{(i,j+1)}_{m}$,
and therefore implicitly constrain variations in the spectral
index.  Wavelet maxima continuously merge in passing from fine to coarse scales,
with no new maxima created at lower resolutions.
This allows us to predict the location of maxima as we go to finer
scales, with the creation of maxima only allowed if the data itself
makes a strong case for it.  In addition, wavelet maxima have 
been demonstrated to be a numerically stable
representation of images, and a direct linear
reconstruction solution given by the amplitude of the maxima
(Mallat and Zhong 1991, Carmona et al. 1998).  A generalization
of the wavelet maxima representation is given by
Basis Peruit (Chen et al. 1999), in which
a decomposition is given by
\begin{equation}
\begin{array}{ll}
T^{(i,0)}_{m}(r) = \sum_{k} \alpha^{(i)}_{m,k} \psi_{k}(r) & 
\mbox{ such that ${\cal S}[\alpha^{(i)}_{m,k}]$ is a minimum}
\end{array}
\end{equation}
where $\psi_{k}$ are the wavelet basis functions, and
${\cal S} = \sum_{k} |\alpha^{(i)}_{m,k}|$ quantifies the sparseness of the
wavelet coefficients.  Basis Persuit has been shown to generate solutions that are
very close to the wavelet maxima representation, demonstrating that
maxima can be understood as optimal representations in a specific sense
for a wide variety of images.

We therefore propose that the relevant variations
in the spectral index to consider are {\it the variations in the
spectral index which increase the likelihood and which are
generated from a sparse representation} according to
\begin{equation}
c_{\nu}[\beta^{(i,j)} - \beta^{(i,j+1)}] = 
\log \left( G^{j} \ast \sum_{k} \alpha^{(i)}_{m,k} \psi_{k}(r) \right)
 - \log T^{(i,j)}_{0} \left( \frac{\nu_{m}}{\nu_{0}}
\right)^{\beta^{(i,j+1)}}
\end{equation}
where the wavelet coefficients should be consistent with the constraint
\begin{equation}
T^{(i,j+1)}_{m} = T^{(i,j+1)}_{0} \left( \frac{\nu_{m}}{\nu_{0}}
\right)^{\beta^{(i,j+1)}(r)} = G^{j+1} \ast \left( 
\sum_{k} \alpha^{(i)}_{m,k} \psi_{k}(r) \right)
\end{equation}
Formally then, we can give a prior for the spectral index as
a marginal process on the ``hidden'' wavelet coefficients according to
(with the content on the right side of the condition bar implied
in the above)
\begin{equation}
p[\beta^{(i,j)}|T^{(i,j)}_{0},T^{(i,j+1)}_{0},\beta^{(i,j+1)}] =
\int d[\alpha^{(i)}_{m,k}]
p[\beta^{(i,j)}|\alpha^{(i)}_{m,k}, \cdots] p[\alpha^{(i)}_{m,k}| \cdots]
\end{equation}
where we can take
\begin{equation}
\begin{array}{l}
- \log p[\beta^{(i,j)}|\alpha^{(i)}_{m,k}, \cdots] \sim \\
\theta \left( c_{\nu}[\beta^{(i,j)} - \beta^{(i,j+1)}] -
\log \left( G^{j} \ast \sum_{k} \alpha^{(i)}_{m,k} \psi_{k}(r) \right)
 + \log T^{(i,j)}_{0} \left( \frac{\nu_{m}}{\nu_{0}}
\right)^{\beta^{(i,j+1)}} \right)^{2}
\end{array}
\end{equation}
and
\begin{equation}
- \log p[\alpha^{(i)}_{m,k}| \cdots] \sim {\cal S} [\alpha^{(i)}_{m,k}]
\end{equation}
To actually compute the marginalization will require
the Metropolis algorithm or Gibbs sampler (Geman and Geman 1984).
We could simply maximize the above, and always replace $\beta$
by the maximum - various approximations will have to be numerically
experimented with to see what works.  

By constraining the emission
of a physical component at each frequency to be given
according to a sparse representation, we have greatly reduced
the effective degrees of freedom.  This approach exploits 
non-Gaussianity, as non-Gaussian features will
typically have long scale-space lifetimes and sparse representations
in a wavelet basis.  The numerical implementation of the above is, needless to say,
much more complicated than the linear deterministic solutions
discussed previously, driving us to stochastic relaxation.  The local
character of the posterior however, gives a conditional probability
structure that enables various degrees of freedom to be adjusted in parallel.

\section{Conclusions}
Bayesian inference does not provide any single method for the
separation of foregrounds and the CMB, but instead is a
framework within which methods can be formulated, and in the process,
explicitly reveal any assumptions made.
The specific method to be used depends on what information
is desired.  Various analysis approaches can be understood
within a Bayesian framework as the ``clean sky'' limit, or the
``high SNR'' limit, or the ``only one suspected foreground component
in these frequencies'' limit, etc.  It seems that a unified view of analysis
would be important, expecially when comparing data returned from different
experiments.  Probably the best way to proceed with actual data
is to attack with any 'reasonable' Bayesian approach imaginable.  With
tools such as a multi-resolution approach and stochastic relaxation,
we can attempt inference within the context of physically
motivated models that address potential complexities in the
foregrounds.  Solutions that are consistent with approaches of varying 
complexity can be considered robust.  However, discrepancies in inferences obtained with
various methods provide an opportunity to learn about the
data itself, and  point the way to needed follow up studies
and future experiments.

\acknowledgements
We thank the late David N. Schramm for continued support and encouragement 
throughout this work.  J. Jewell thanks Yali Amit and Charles Anderson for very helpful
discussions on Gibbs random fields and stochastic algorithms, and
Edith Huang for computational support on the Cray T3D at the Jet Propulsion Lab.
Support for J. Jewell was provided under a NASA Graduate Student Research Program
Fellowship at the Jet Propulsion Lab.

\end{document}